\documentclass[ showpacs,preprint]{revtex4}
\usepackage[dvips]{epsfig,graphicx}
\usepackage{float}
\usepackage[dvipdfm,colorlinks=true, citecolor=blue, urlcolor=blue ]{hyperref}
\usepackage[mathscr]{euscript}
\usepackage{amsfonts}
  
\def\be{\begin{equation}}
\def\ee{\end{equation}}
\def\la{\langle}
\def\ra{\rangle}

\begin{document}
\title{
Critical Casimir Forces for Films with
Bulk Ordering Fields
}

\author{O. A. Vasilyev}
\affiliation{Max-Planck-Institut f{\"u}r Intelligente Systeme,
  Heisenbergstra{\ss}e~3, D-70569 Stuttgart, Germany}
\affiliation{ IV. Institut f{\"u}r Theoretische Physik,
  Universit{\"a}t Stuttgart,  Pfaffenwaldring 57, D-70569 Stuttgart, Germany}
\author{S. Dietrich}
\affiliation{Max-Planck-Institut f{\"u}r Intelligente Systeme,
  Heisenbergstra{\ss}e~3, D-70569 Stuttgart, Germany}
\affiliation{ IV. Institut f{\"u}r Theoretische Physik,
  Universit{\"a}t Stuttgart,  Pfaffenwaldring 57, D-70569 Stuttgart, Germany}

\date{\today}

\begin{abstract}
The confinement of long-ranged critical fluctuations 
in the vicinity of  second-order phase transitions
in fluids generates   critical Casimir forces acting
on confining surfaces or among particles immersed in a critical solvent.
This is realized in binary liquid mixtures close to  their
consolute point $T_{c}$ which 
 belong to the universality class of the Ising model.
The deviation of the difference of the chemical potentials of the two 
species of the mixture from its value at criticality corresponds to  
 the bulk magnetic filed of the Ising model.
By  using Monte Carlo simulations
 for this latter representative of the corresponding universality class
   we compute the critical Casimir force as a
function of the bulk ordering field  at the critical temperature $T=T_{c}$.
We use a coupling parameter scheme for the computation of the 
underlying free 
energy differences and an energy-magnetization integration  
 method for computing  the bulk free energy density
 which is a necessary ingredient.
By taking into account finite-size corrections,
for various types of boundary conditions
 we determine the universal Casimir force 
scaling function as a function of the  scaling variable
associated with the bulk field.
Our numerical data are compared with 
analytic results obtained from mean-field theory.

\end{abstract}


\maketitle

In the vicinity of  second-order
phase transitions  long-ranged fluctuations
of the corresponding order parameter arise.  Fisher and de~Gennes 
pointed out that in fluids the spatial confinement of such fluctuations 
produces  effective forces  acting on the confining  surfaces~\cite{FdG}.
In view of certain similarities 
 with the electromagnetic Casimir 
effect~\cite{Casimir,KG}, in which such
 forces are induced by the quantum fluctuations 
 of the electromagnetic field,
these forces in critically fluctuating media are called 
{\it critical Casimir forces}~(CCF)~\cite{Krech,BDT,Gambassi}. 
In line with the finite size scaling concept~\cite{Barber,Privman}  
 CCF are characterized by universal scaling functions depending on the ratio
of the distance between the confining surfaces  and the bulk 
correlation length $\xi$, which diverges upon approaching the critical point  
$T_{c}$~\cite{Krech,BDT,Gambassi}.
The scaling function depends on the bulk universality class
and on the type of boundary conditions (BC) for the order parameter.
For classical binary liquids mixtures, which belong to the 
Ising bulk universality class, CCF have been measured experimentally both
indirectly via their influence on wetting films~\cite{Fukuto}
and directly by monitoring a colloidal particle
near a wall and immersed in a critical solvent~\cite{nature, PRE1}.
There is excellent  agreement 
between these experimental data and the corresponding theoretical 
results~\cite{EPL,PRE,sp}.  
  
\begin{figure}
\mbox{\includegraphics[width=0.175\textwidth]{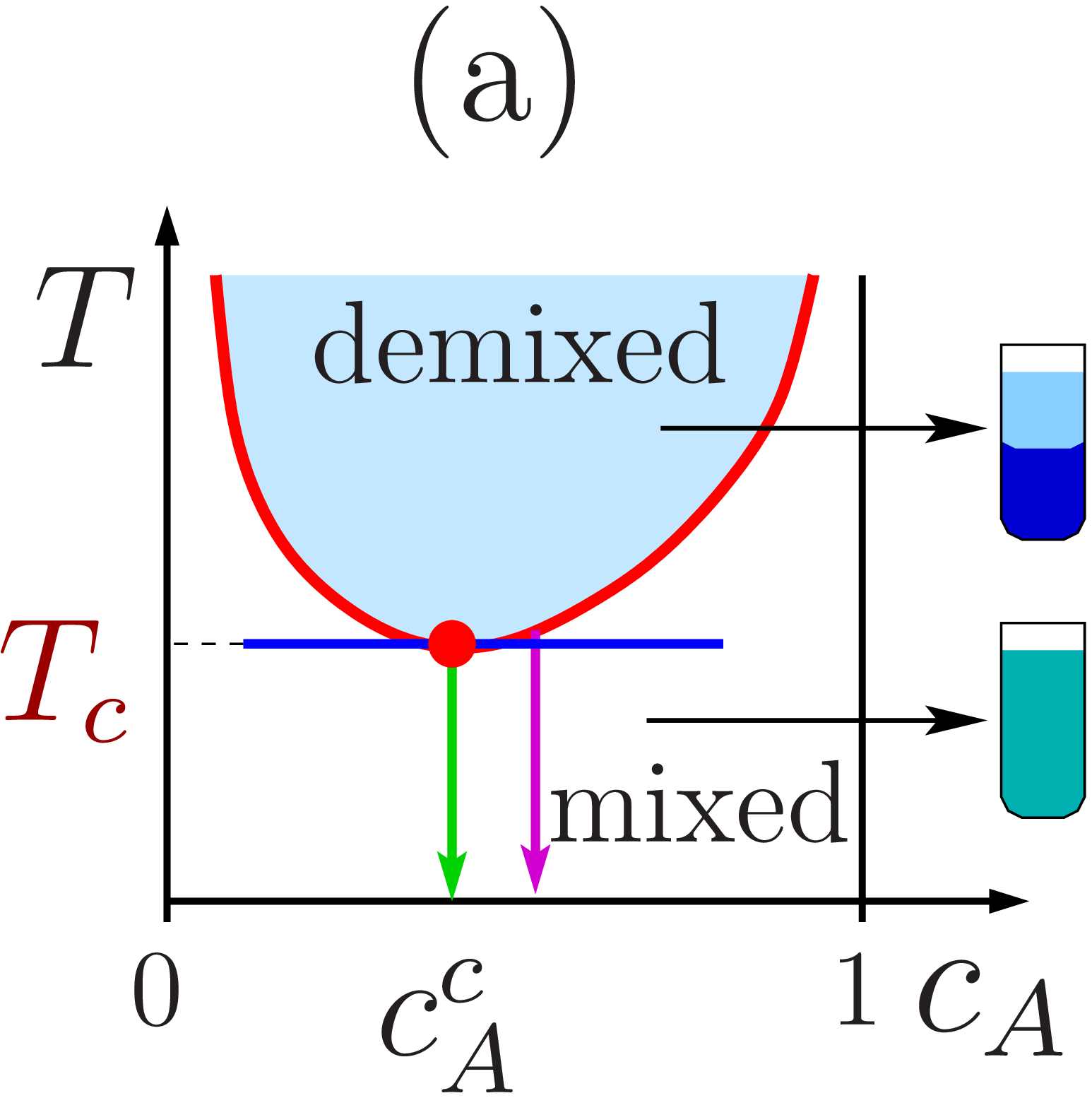}
\includegraphics[width=0.135\textwidth]{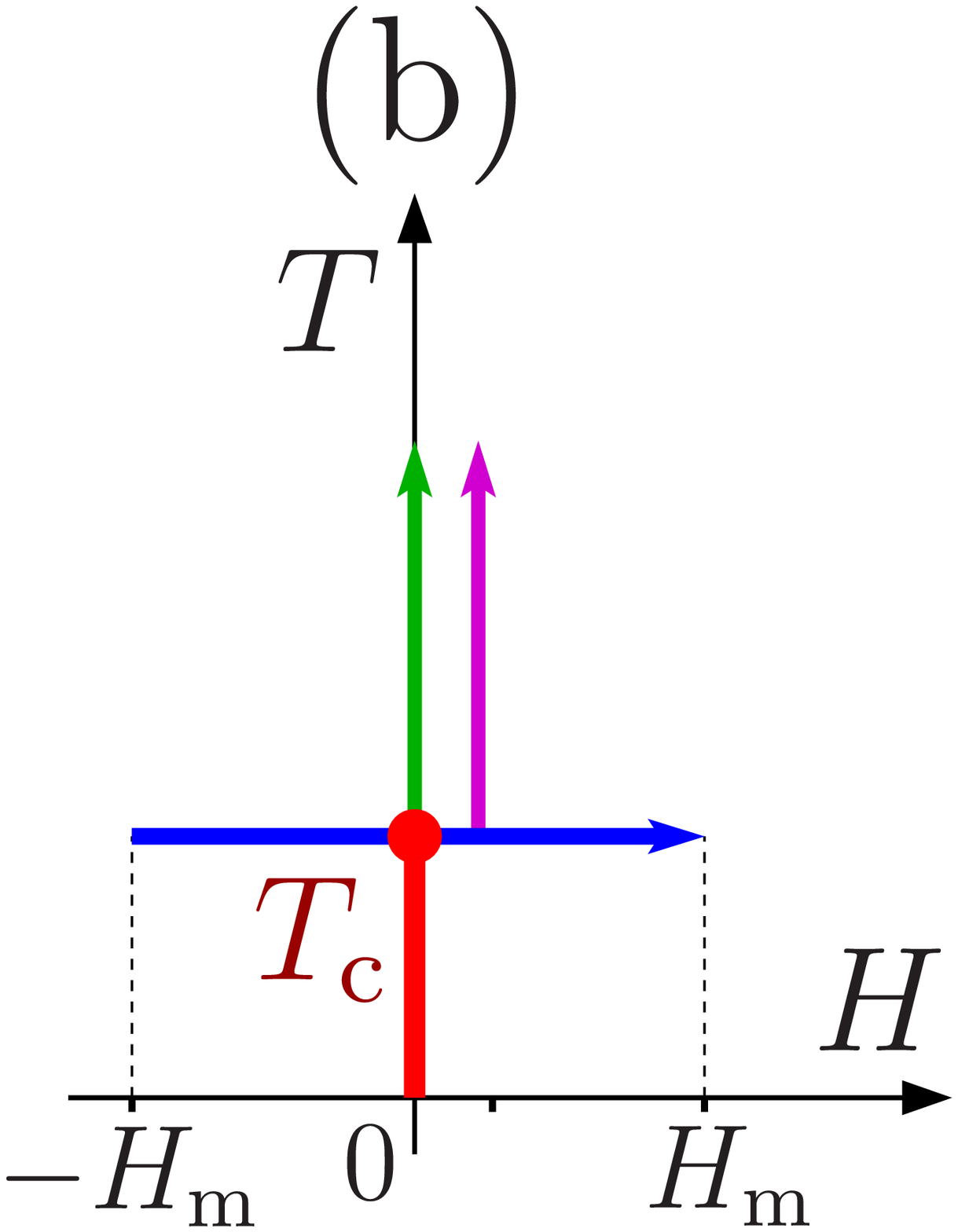}
\includegraphics[width=0.16\textwidth]{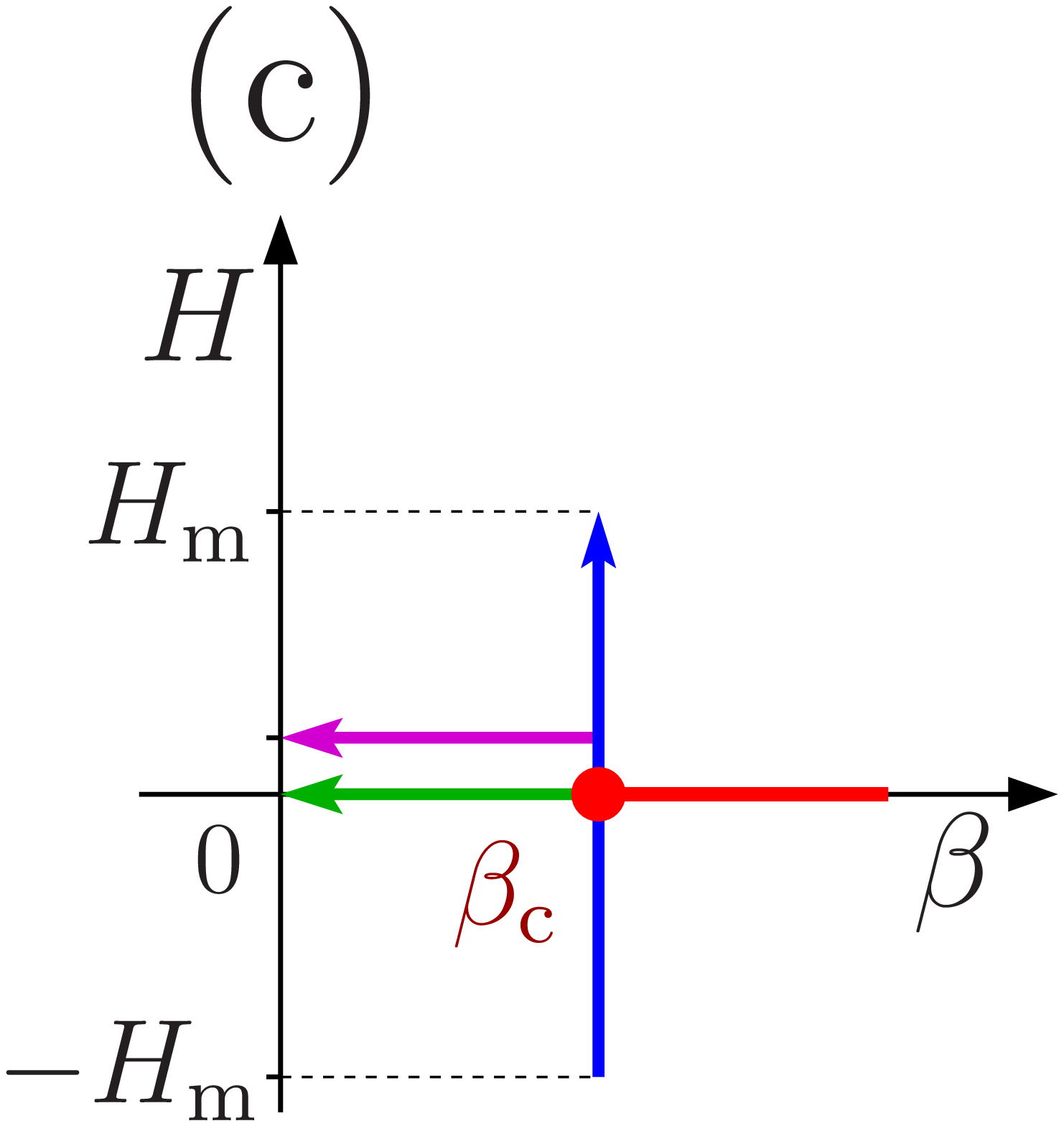}}
\caption{%
 (a) Schematic phase diagram of demixing in binary liquid mixtures
 with a lower critical point at $(T=T_{c},c_{A}=c_{A}^{c})$
 where $T$ is the temperature and $c_{A}$ is the concentration 
 of one of the two species of the mixture. The green, magenta, and blue
 full lines indicate three distinct thermodynamic paths.
 (b) Phase diagram  of the Ising model in the $(H,T)$
 plane where $H$ is the bulk field. Note that two-phase coexistence 
 for $T \ge T_{c}$ in (a) corresponds to $(H=0,T\le T_{c})$ in (b). 
 The isotherm runs  in the interval $|H| \le H_{\mathrm m}$  (see main text)
 and the magenta path corresponds to $(T>T_{c}, H=0.17 H_{\mathrm{m}})$.
 (c) Phase diagram and corresponding paths in the 
 $(\beta=1/(k_{\mathrm B}T),H)$ plane with $\beta_{c}=1/(k_{\mathrm B}T_{c})$.
 }
\label{fig:fig1}
\end{figure}

  Figure~\ref{fig:fig1}(a) shows the schematic bulk phase diagram for the type
  of binary liquid mixtures (such as water-lutidine)
  used in these experiments~\cite{Fukuto,nature,PRE1}; they exhibit
a lower critical point $(T_{c},c_{A}^{c})$
where $c_{A}$ denotes the concentration of one of the 
two components $A$ and $B$ (e.g., lutidine) of the mixture.  
 Long-ranged fluctuations of the order parameter $\psi \sim c_{A}-c_{A}^{c}$
  arise upon approaching this point either along an iso-concentration 
  $c_{A}=c_{A}^{c}$ path or along an isotherm $T=T_{c}$ (or any other direction).
  The phase diagram for the corresponding
  Ising model is shown in Fig.~\ref{fig:fig1}(b).
 The bulk magnetic
 field $H$ plays the role of 
 $\mu_{A}-\mu_{B}-(\mu_{A}-\mu_{B})_{c}$
 where $\mu_{A,B}$ are the chemical potentials of the two species 
 of the fluid.
 Together with the reduced temperature $t=(T-T_{c})/T_{c}$
  this difference determines the order parameter  $c_{A}-c_{A}^{c}$.
 The scaling functions of CCF depend strongly on the BC.
Generically, one of the two species of the binary mixture is preferentially
adsorbed at a confining wall which within the Ising model corresponds to the
 presence of a (strong) surface field, denoted as $(+)$
 or $(-)$ BC. If the surface is neutral with respect to the two species 
one is lead to Dirichlet BC (denoted as (O)~)~\cite{Diehl}.
 For the Ising universality class and in the presence
 of surface fields the variation of the CCF upon varying the BC has been 
 studied experimentally~\cite{Nellen}, theoretically~\cite{Mohry},
  and numerically~\cite{Hass,Surf}. One finds a continuous crossover
  between attractive CCF for $(+,+)$ BC and repulsive ones for $(+,-)$ BC.
  There is experimental evidence that CCF do not only depend
  sensitively   on temperature but also on $c_{A}$~\cite{BE,Nellen1}.
 However, whereas there is by now rather reliable theoretical knowledge 
 concerning the temperature dependence of CCF~\cite{EPL,PRE,Hass,Surf},
 there are only a few studies of their concentration dependence;
  they are either pure mean-field studies~\cite{SHD}
 or scaling-theory enhanced mean-field studies~\cite{Mohry1,BCPP1,BCPP2}.
 For spatial dimension $d=2$ the CCF 
in the presence of a bulk magnetic field have been  studied 
in detail in Refs.~\cite{DMC,MDC,MDE}.

 In particular, for spatial dimension $d=3$
 there are no simulation data available concerning the dependence of the CCF on the 
 bulk magnetic field within the Ising universality class.
 The present study closes this gap and provides insight into the scaling
 behavior of CCF in the full
 neighborhood of the critical point for four sets of BC: $(+,+)$,
 $(-,+)$, $(O,+)$, and $(O,O)$.  
 
We consider a simple
cubic lattice with  lattice spacing $a$. (On the lattice all lengths are measured 
 in units of $a$ and thus are dimensionless.)
The lattice sites form  a slab 
 $L_{x} \times L_{y} \times L_{z}$
with $L_{x}=L_{y} = 6 L_{z}$ and with a cross-section $A=L_x \times L_y$. 
There are  periodic BC along  the  $x$ and $y$ axes.  
In our study we have carried out  simulations for 
 $L_{z}=10,15$, and 20.

Each lattice site $i=( 
1 \le x \le L_{x}, 1 \le y \le L_{y}, 1 \le z \le L_{z})$  is occupied by
a spin $s_{i}=\pm 1$.  
The Hamiltonian of the Ising model with bulk ($H$) and surface fields
($H_{1}^{\pm}$, acting on the bottom [$-$] and the top [$+$]
layers $z=1,L_{z}$, respectively)   is
\be
\label{eq:Ham}
{\cal H} = -  \sum_{\la {\rm ij} \ra}  s_{i}  s_{j}
-H\sum_{ k }s_{k} 
-H_{1}^{-} \sum_{\la \mathrm{ bot.} \ra}s_{j}-
H_{1}^{+}\sum_{\la \mathrm{ top} \ra}s_{j}. 
\ee
Here and in the following the energies and fields are 
measured in units of the spin-spin interaction constant $J$.
The sum $\la {\rm ij} \ra$ is taken  over all nearest-neighbor pairs
of sites on the lattice and  the sum over $ k$ runs
 over all spins. The four types of BC which we study correspond to 
$(H_{1}^{-},H_{1}^{+})=(+\infty,+\infty) \equiv (+,+)$, 
$(-\infty,+\infty) \equiv (-,+)$, 
$(0,+\infty) \equiv (O,+)$, and
$(0,0) \equiv (O,O)$.
In practice, we use   surface fields
which  are  finite but strong enough 
to observe  saturation of results and thus  mimic the action 
of  infinite surface fields~\cite{Surf}. 
Finite surface fields give rise to a dependence on the scaling variables 
$H_{1}^{\pm}L_{z}^{\Delta_{1}/\nu}$~\cite{Diehl};
 we use  $H_{1}L_{z}^{\Delta_{1}/\nu}=+100$ and $H_{1} L_{z}^{\Delta_{1}/\nu}=- 100$
 instead of $+\infty$ and $-\infty$,  respectively.
Here   $\nu=0.6301(4)$~\cite{PV} is the critical exponent
of the bulk correlation length 
$\xi^{\pm}_{t}\left(
t=\frac{T-T_{c_{\mbox{\rule{0pt}{2.5pt}} }}}
{T_{c}} \to \pm 0, H=0\right)=\xi_{t,0}^{\pm}|t|^{-\nu}$,
  and  $\Delta_{1}=0.46(2)$~\cite{GZ}
is the so-called critical surface gap exponent. 
For these large values for  $H_{1}^{\pm}$ and $A$
the  system depends de facto only on the  three parameters
$\beta=1/(k_{\mathrm{B}}T)$, $H$, and $L_{z}$.
The critical value of  $\beta$ is 
$\beta_c=1/(k_{B}T_{c})=0.2216544(3)$~\cite{RZW}.

According to  finite-size scaling theory~\cite{fisher_nakanishi}, 
for given  BC and number of layers $L_{z}$
 the thermodynamic  state of the system is
characterized  by two scaling variables:
$(L_{z}/\xi_{t},H L_{z}^{\Delta /\nu})$, where
$ H L_{z}^{\Delta /\nu}$ is the bulk magnetic field
scaling variable with   $\Delta=1.5637(14)$~\cite{PV}.

For large values of $A$, the total free energy
$F(\beta,H ,L_z)$ of the  film  can be written as
$ F(\beta,H,L_z) =
A \beta^{-1} [L_{z} f^{\mathrm{b}}(\beta,H)+
f^{\mathrm{ex}}(\beta,H,L_z)]$.
Here  $f^{\mathrm{b}}(\beta,H)$ 
 is the bulk free energy density per $k_{\mathrm B}T$ 
 of the macroscopic system
 at a given temperature and bulk magnetic field.
The  excess free energy $f^{\mathrm{ex}}$ per area 
gives rise to the critical  Casimir force  $f_{\mathrm{C}}$ 
 in units of $k_{\mathrm B} T$ and $A$:
$f_{\mathrm{C}}(\beta,H,L_{z})
\equiv - \partial f^{\mathrm{ex}}(\beta,H,L_{z})/\partial L_{z}$.
For given BC, on a lattice (we denote lattice quantities by symbols 
with a ``hat'' $\,\hat{}\,$~) 
we replace the derivative  by the finite difference 
\be 
\label{eq:force}
\hat f_{\mathrm{C}}^{(BC)}\left(\beta,H,L\right)
:= - \frac{\beta \Delta \hat F^{(BC)}(\beta,H,L_{z},A)}{A}
+  \hat f^{\mathrm{b}}(\beta,H)\,,
\ee
where  
$\Delta \hat F^{(BC)}(\beta,H,L_{z},A)= 
\hat F^{(BC)}(\beta,H,L_{z},A)-
\hat F^{(BC)}(\beta,H,L_{z}-1,A)$.
Here we express the CCF in terms of the film thickness $L:=L_{z}-\frac{1}{2}$
which is a half-integer quantity.

\begin{figure}
\includegraphics[width=0.45\textwidth]{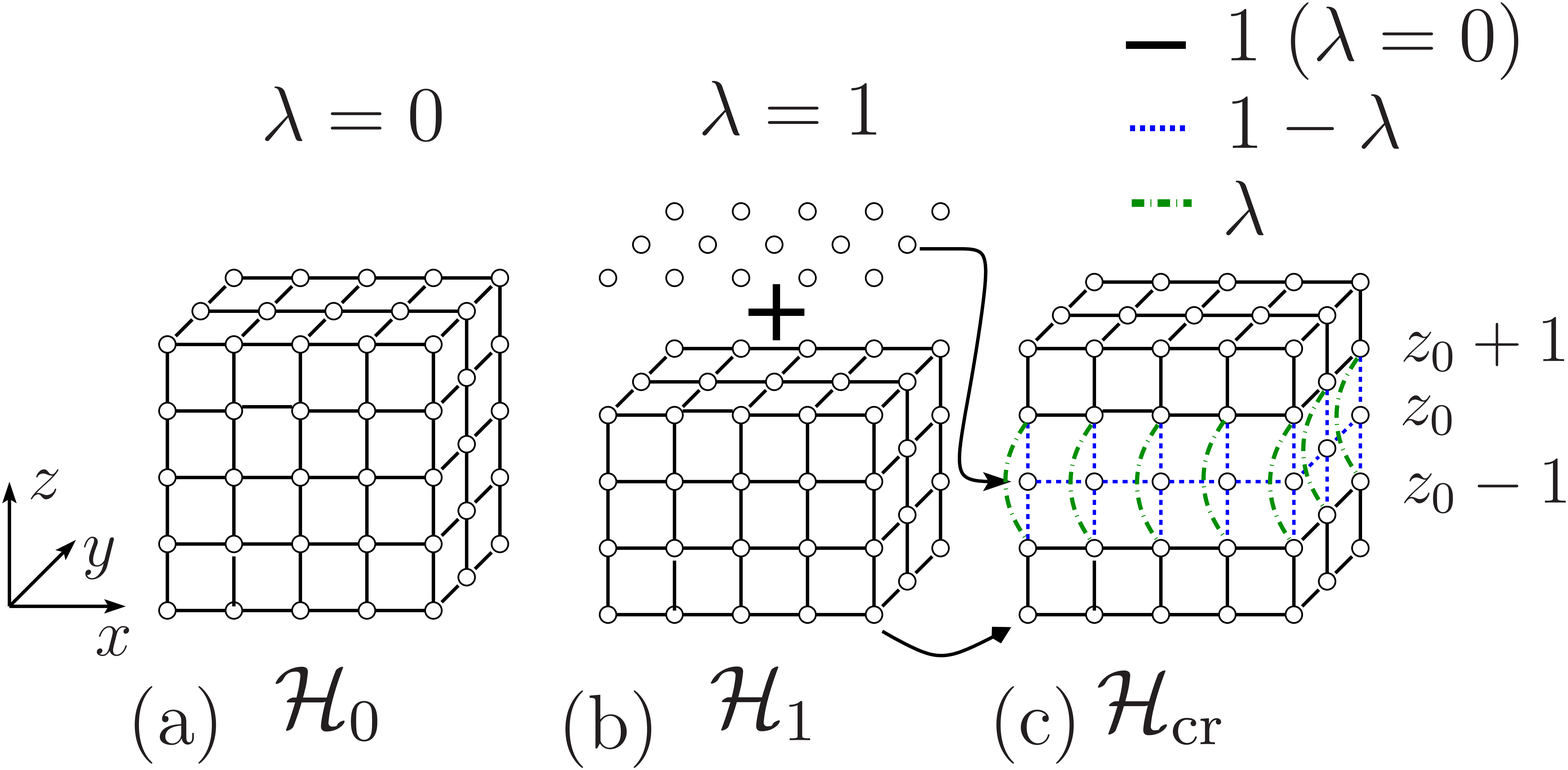}
\caption{%
Arrangement of bonds for determining  the free energy
difference between systems
with Hamiltonian ${\cal H}_{0}$ and $L_{z}$
layers (a)  and with Hamiltonian ${\cal H}_{1}$
and $L_{z}-1$ layers plus  $A=L_{x}\times L_{y}$
 isolated spins (b).
The crossover Hamiltonian 
${\cal H}_{\rm cr}={\cal H}_{0}+\lambda ({\cal H}_{1}-{\cal H}_{0})$
 interpolates between  ${\cal H}_0$ and ${\cal H}_1$
upon changing $\lambda$  from 0 to 1 (c).
}
\label{fig:fig2}
\end{figure}
In accordance with eq.~(\ref{eq:force}), 
we determine the film free energy difference $\Delta  \hat F^{(BC)}$
and the bulk free energy $\hat f^{\mathrm{b}}$ per spin and per $k_{\mathrm B}T$
as functions of the bulk magnetic field $H$ at $T_{c}$.
To this end we use the  coupling parameter approach 
(see  Refs.~\cite{Mon,PRE,Surf}).    
In this context ${\cal H}_{0}$ denotes the Hamiltonian of the system with $L_{z}$ layers 
 [Fig.~\ref{fig:fig2}(a)] and 
${\cal H}_{1}$ is the Hamiltonian of the system with
$L_{z}-1$ layers plus a layer of $A=L_{x}\times L_{y}$ isolated spins
 [Fig.~\ref{fig:fig2}(b)] which keeps the number of spins
  in the system constant.
 We introduce the 
 {\it cr}ossover Hamiltonian  
$ {\cal H}_{\rm cr}(\lambda)=
{\cal H}_{0}+\lambda \Delta {\cal H}$,
with  $\Delta{\cal H}= {\cal H}_{1}-{\cal H}_{0}$,
which interpolates between ${\cal H}_{0}$ and ${\cal H}_{1}$,
 upon changing the  coupling parameter 
 $\lambda$  from 0 to~1, by  suitably varying
 certain interaction constants as $\lambda$ and $1-\lambda$ 
(see Fig.~\ref{fig:fig2}(c))
 for a selected layer at height $z_{0}=L_{z}/2\;[(L_{z}+1)/2]$ for 
 even~[odd] values of $L_{z}$.
 The free energy difference between these two systems
 is $\Delta F=\int_{0}^{1} F'_{\rm cr}(\lambda) {\rm d}\lambda
 =\int_{0}^{1}\la \Delta {\cal H}\ra_{\rm  cr}(\lambda) {\rm d}\lambda$
 where the free energy $F_{\rm cr}(\lambda)$ corresponds to ${\cal H}_{\rm cr}(\lambda)$
and its derivative $F'_{\rm cr}(\lambda)=
\frac{\mathrm d}{{\mathrm d}\lambda}F(\lambda)=\la \Delta {\cal H}\ra_{\rm cr}(\lambda) $
 takes the form of the canonical ensemble average
  $\la \ldots \ra_{\rm cr}(\lambda)$
  taken with $\exp(-\beta {\cal H_{\rm cr}})$ 
of the energy difference $\Delta {\cal H}$.
  We have determined
the ensemble averages $\la \Delta {\cal H}\ra_{\rm  cr}(\lambda)$ 
via MC simulations for
 $N_{\lambda}=21$
 different values of $\lambda_{k}=\frac{k}{N_{\lambda}-1}$
 ($k=0,\dots,N_{\lambda}-1$) by using the 
 hybrid MC method  with a mixture of Wolff 
and  Metropolis algorithms.
For the computation of the thermal 
average we have used $5 \times 10^{5}$ MC steps [$ 10^{6}$ for $(O,O)$ BC].
Based on $N_{\lambda}$ points we have performed the
  numerical integration over $\lambda$
  by  using Simpson's rule.
Accordingly, the free energy difference appearing in eq.~(\ref{eq:force})
is given by 
\begin{eqnarray}
\Delta \hat F^{(BC)}(\beta,H,L,A) =\\ \nonumber
 -  \int_{0}^{1}  \langle \Delta
{\cal H} \rangle_{\rm cr}(\lambda) {\rm d}\lambda
-A\beta^{-1} \ln[2 \cosh(\beta H) ],
\label{eq:df}
\end{eqnarray}
where the last term corresponds to the free energy of $A$ isolated spins.

Once $\Delta \hat F^{(BC)}(\beta,H,L,A)$
 has been computed, one still has to separate off 
$\hat f^{\mathrm{b}}(\beta_{c},H)$ 
from it [see eq.~(\ref{eq:force})] in order to obtain the
Casimir force.   
In the absence of the bulk magnetic field $H$ the bulk free energy 
can be determined  via  temperature integration~\cite{hucht,hasen1,hasen2}. 
We extend this method to the case $H \ne 0$.
To this end, as in Ref.~\cite{Surf}, we determine the free energy density
for a cube of volume  $L_{\rm cube}^{3}=128^{3}$
with periodic BC in all directions. We consider  this value
as the desired bulk free energy density (per $k_{\mathrm{B}}T$): 
 $  \hat f^{\mathrm{b}}(\beta,H) \simeq
 \hat f^{\mathrm{cube}}(\beta,H,L_{\mathrm{cube}}=128)$.
 In order to obtain $\hat f^{\rm cube}$ we have integrated
 the appropriate  combination 
  $ E (\beta',H) - H  M (\beta',H)  $
 of the energy and the magnetization:
\begin{eqnarray}
\label{eq:fb2}
 \hat f^{\mathrm{cube}}(\beta,H)=
-\ln(2)+\\ \nonumber
L_{\rm cube}^{-3}
\int \limits_{0}^{\beta} \left[  E (\beta',H)
-H  M (\beta',H) \right]{\rm d }\beta',
\end{eqnarray}
where 
$  E (\beta,H)=-\la \sum \limits_{ \la i,j \ra} 
s_{i}s_{j} \ra_{\mathcal{H}(H)}$ and 
$  M (\beta,H)=\la \sum \limits_{k}
s_{k} \ra_{\mathcal{H}(H)}$
are the energy and  the magnetization, respectively,
of a system at an inverse temperature $\beta$ and
with  a bulk magnetic  field $H$.
\begin{figure*}[t]
\mbox{\includegraphics[width=0.45\textwidth]{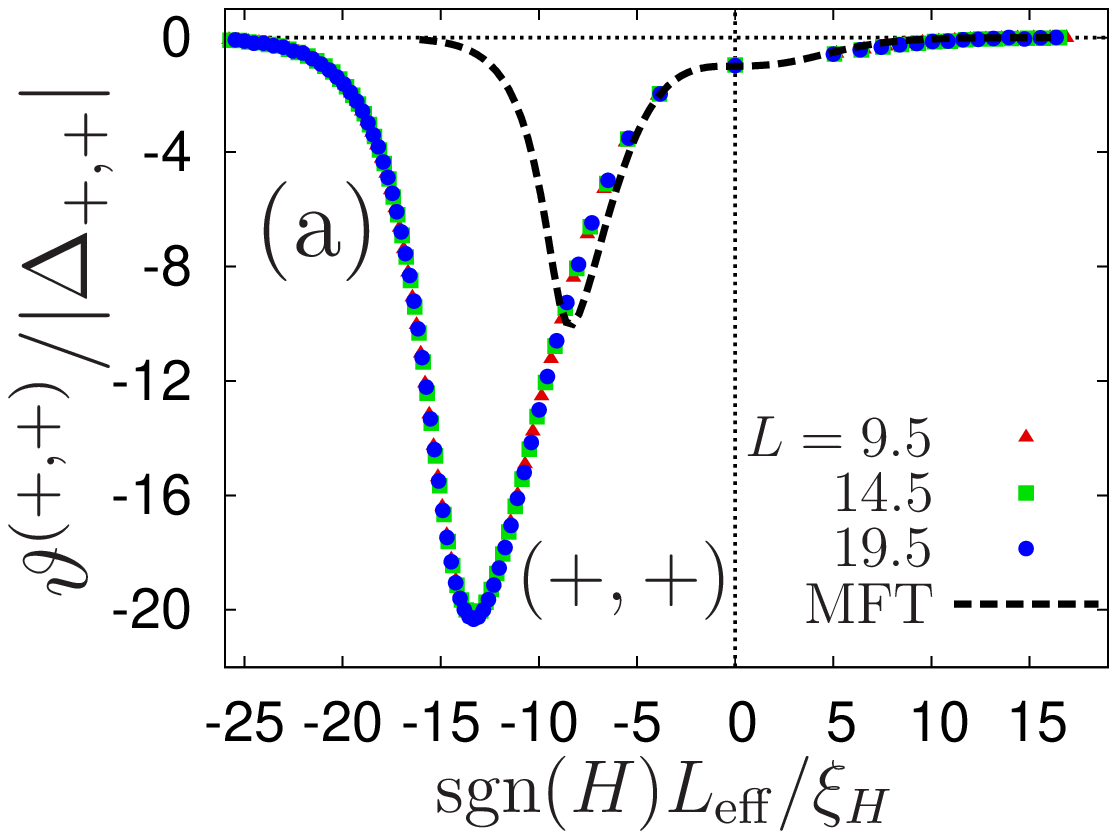}
\includegraphics[width=0.45\textwidth]{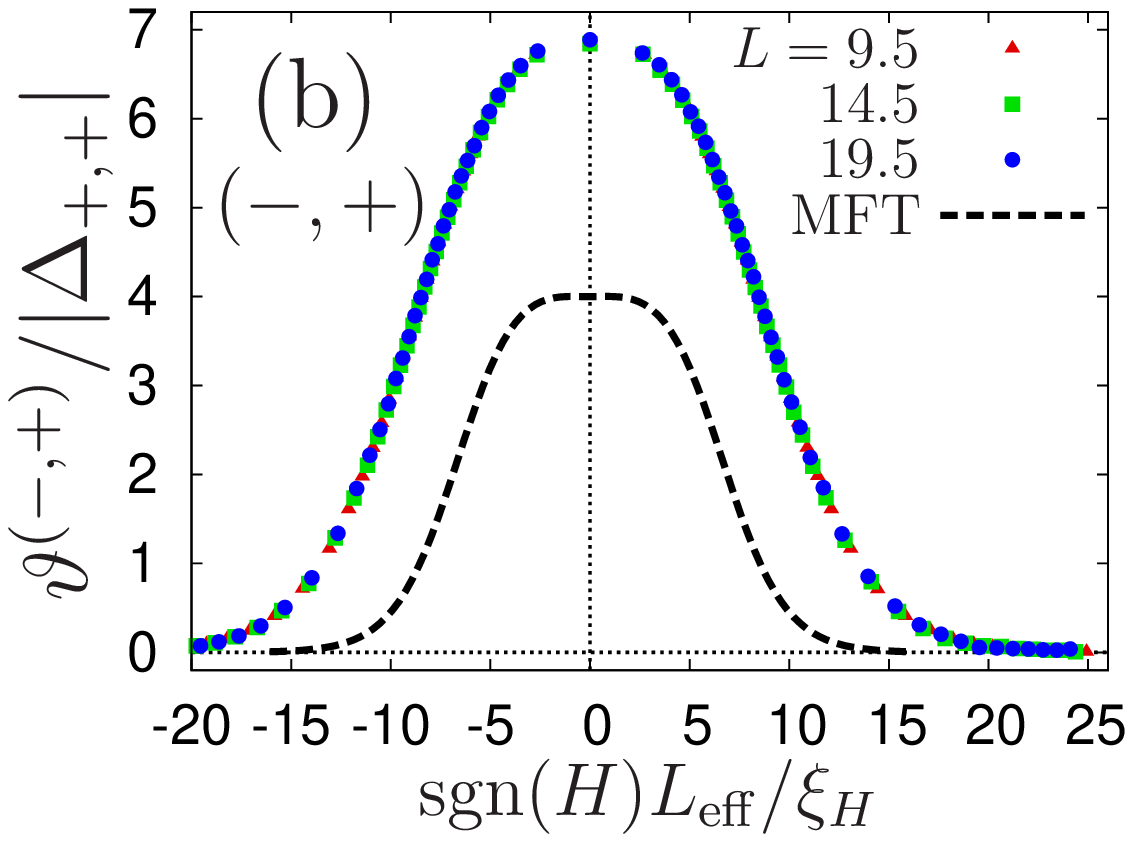}}\\
\mbox{\includegraphics[width=0.45\textwidth]{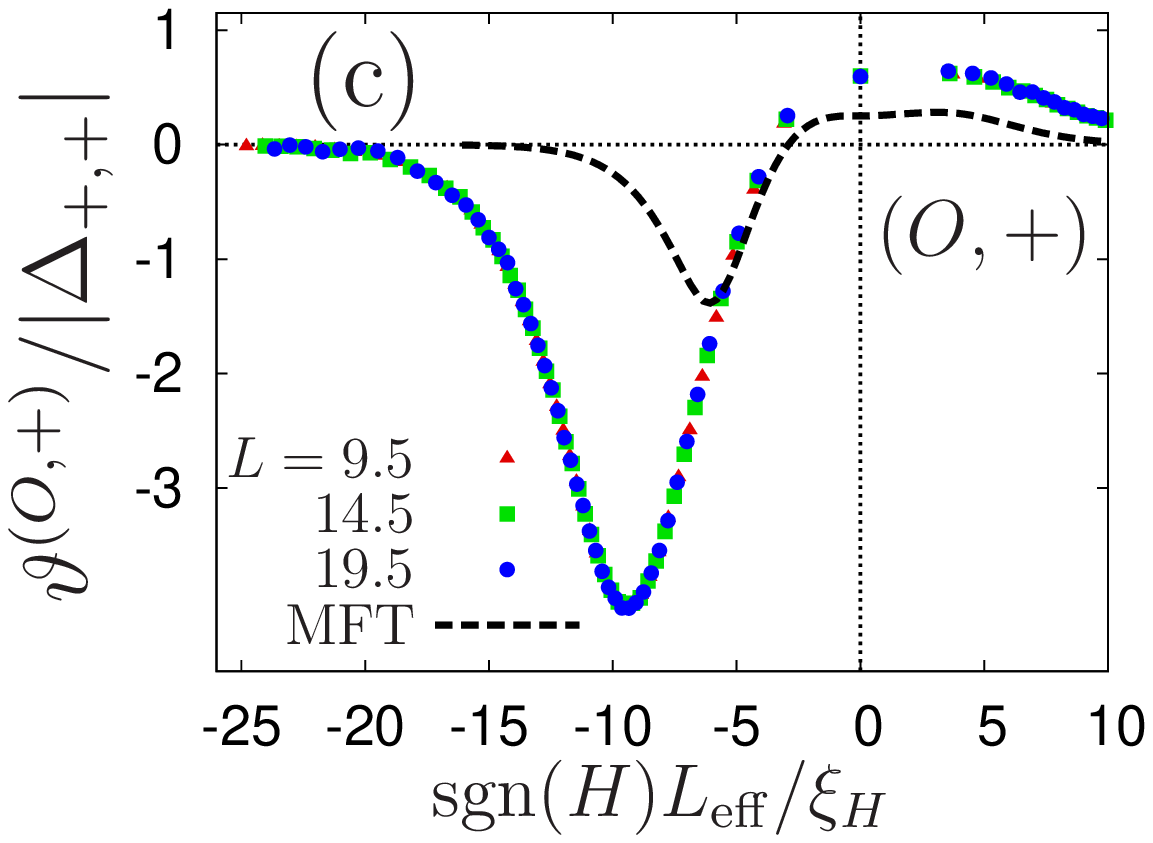}
\includegraphics[width=0.45\textwidth]{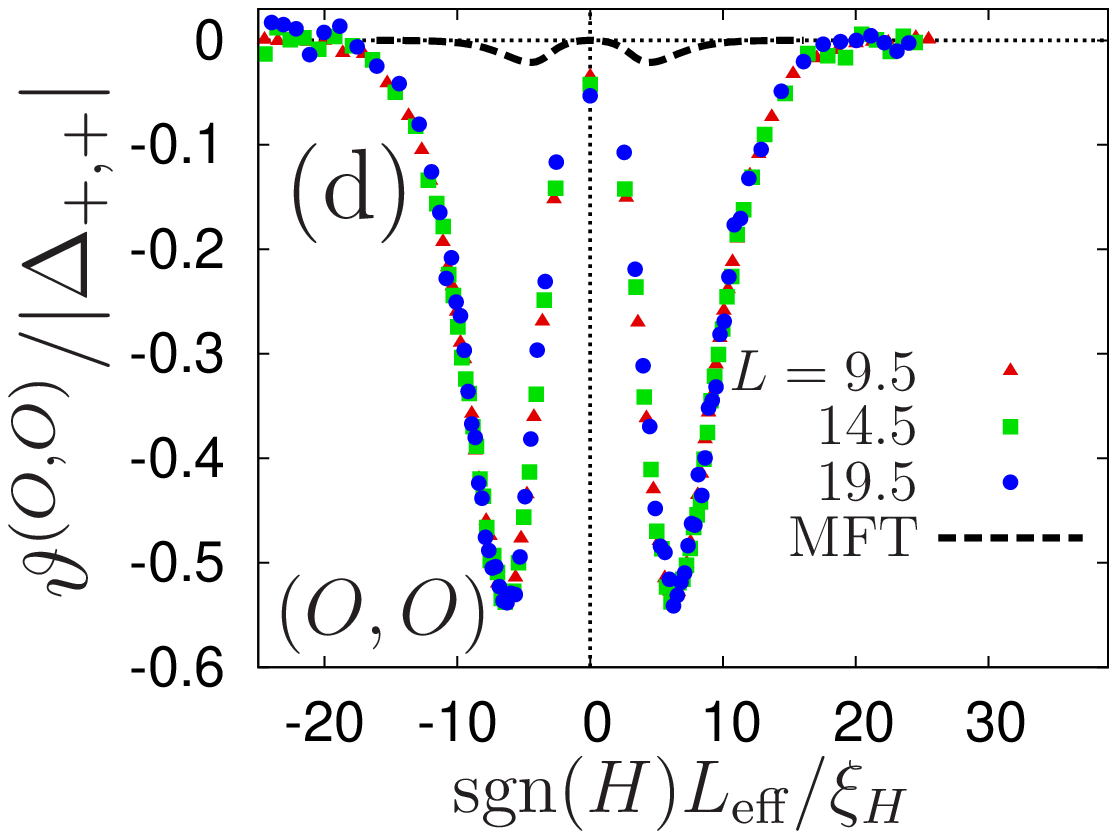}}
\caption{%
 The MC data points show the universal scaling functions 
(eq.~(\ref{eq:sc})) $\vartheta^{(BC)}/|\Delta_{++}|$ for $d=3$,
$T=T_{c}$,  and $L=L_{\mathrm{eff}}-\delta L=9.5,14.5$, 
 and $19.5$ (Table~\ref{tab:dl}) normalized by the critical  Casimir amplitude 
 $\Delta_{++}(d=3) \simeq  -0.75(6)$ 
 ($T=T_{c}$ and $H=0$)~\cite{PRE},
 as functions of the scaling variable ${\mathrm{sgn}}(H) L_{\mathrm{eff}}/\xi_{H}$ 
  for four BC: (a)~$(+,+)$; (b)~$(-,+)$; (c)~$(O,+)$; (d)~$(O,O)$.
  The dashed lines show the corresponding normalized
  (by $\tilde \Delta_{++}(d=4)$)
  universal scaling functions in $d=4$, as obtained within MFT and as function
  of ${\mathrm{sgn}}(\tilde H) \tilde L/\xi_{ \tilde H}$. The MFT expressions 
  for $\tilde \vartheta$
   carry, inter alia, an undetermined prefactor $g^{-1/2}$.
   This dependence on $g$ drops out upon choosing the above normalization,
   rendering a universal ratio in $d=4$. Accordingly, in (a)
   both the MC data and the MFT results attain the value 1 at the origin
   and in  (b) the MFT result attains the value 4 there.
   In (d) the MFT result has a zero at the origin whereas the MC data are
   slightly nonzero there. The results in (b) and (d) are symmetric around the origin.
   Note the different scales of the axes.
}
\label{fig:thetad}
\end{figure*}
 ${\cal H}(H)$ is given by eq.~(\ref{eq:Ham}) with $H_{1}^{\pm}=0$.  
Knowing the free energy density $\hat f^{\mathrm{b}}(\beta,H_{0})$
at a certain value  $H_{0}$ of the bulk magnetic field 
one can compute the bulk free energy density for an arbitrary 
value of the magnetic field $H$
via integration:
\be
\label{eq:fb3}
 \hat f^{\mathrm{b}}(\beta,H)
= \hat f^{\mathrm{b}}(\beta,H_{0})-\beta
L_{\rm cube}^{-3}
\int \limits_{H_{0}}^{H} 
 M (\beta,H')  {\rm d }H'.
\ee
By using eqs.~(\ref{eq:fb2}) and~(\ref{eq:fb3}) we have performed
 numerical integrations along the three paths shown in Fig.~\ref{fig:fig1}(c):
  $(\beta,H=0)$~[green], $(\beta,H=0.1)$~[magenta], 
  and $(\beta=\beta_{c},H)~$[blue].
We have employed a histogram reweighting method~\cite{FS,LB} for improving
the accuracy of the numerical integration. Accordingly, 
for 165 points of $\beta$ in the interval $[0,\beta_{c}]$
we have computed histograms
(averaged over $10^{6}$ MC steps) of the quantities 
$ E (\beta,H=0)$ and 
$ E (\beta,H=0.1) -0.1  M (\beta,H=0.1)$.
 We have also computed the histogram of $M (\beta_{c},H)$ 
 for 256 points  of the bulk field $H$ in the interval 
$0\le H \le H_{\mathrm m}$
where $H_{\mathrm m}=0.59$ (see Figs.~\ref{fig:fig1}(b) and (c)).
For negative values of $H$ we have used the symmetry relation
$f^{\mathrm{b}}(\beta,-H)=f^{\mathrm{b}}(\beta,H)$.
 In a second step we have  performed numerical integration along these trajectories
 using histogram reweighting with the trapezoid rule using $10^{5}$ points.
Integrating along the green line $(\beta,H=0)$ we have obtained the critical  value
$ \hat f^{\mathrm{b}}(\beta_{c},H=0)=-0.77785038(36)$
whereas  sequentially integrating along the magenta 
 $(\beta,H=0.1)$ and blue  $(\beta_{c},H)$ line,
 we have obtained the value 
$ \hat f^{\mathrm{b}}(\beta_{c},H=0)=-0.77784921(60)$
which de facto  coincides within the numerical accuracy
 with the former value.
 To the best of our knowledge, the dependence of the 
bulk free energy $ \hat f_{\mathrm{b}}$ of the $d=3$ Ising model
as a function of the bulk magnetic field $H$ is not yet 
available and the present analysis closes this gap.
Finally, we  combine the results for the bulk free energy 
$ \hat f^{\mathrm{b}}(\beta_{c},H)$
with the corresponding ones  for the free energy difference
 $\Delta \hat F^{(BC)}(\beta,H,L,A)$ leading to the critical Casimir force
\begin{eqnarray}
\hat f_{\mathrm C}^{(BC)}(\beta_{c},H,L)= 
\beta_{c} A^{-1} \int_{0}^{1}  \langle \Delta
{\cal H} \rangle_{\rm cr}(\lambda) {\rm d}\lambda
\\ \nonumber
+ \hat f^{\mathrm{b}}(\beta_{c},H)+
\ln[2\cosh(\beta_{c} H)].
\end{eqnarray}
The numerical accuracy of $\hat f_{\mathrm C}^{(BC)}$ is determined
in a standard way by  subdividing the numerical results into 10 series.

\begin{figure*}[t]
\mbox{\includegraphics[width=0.45\textwidth]{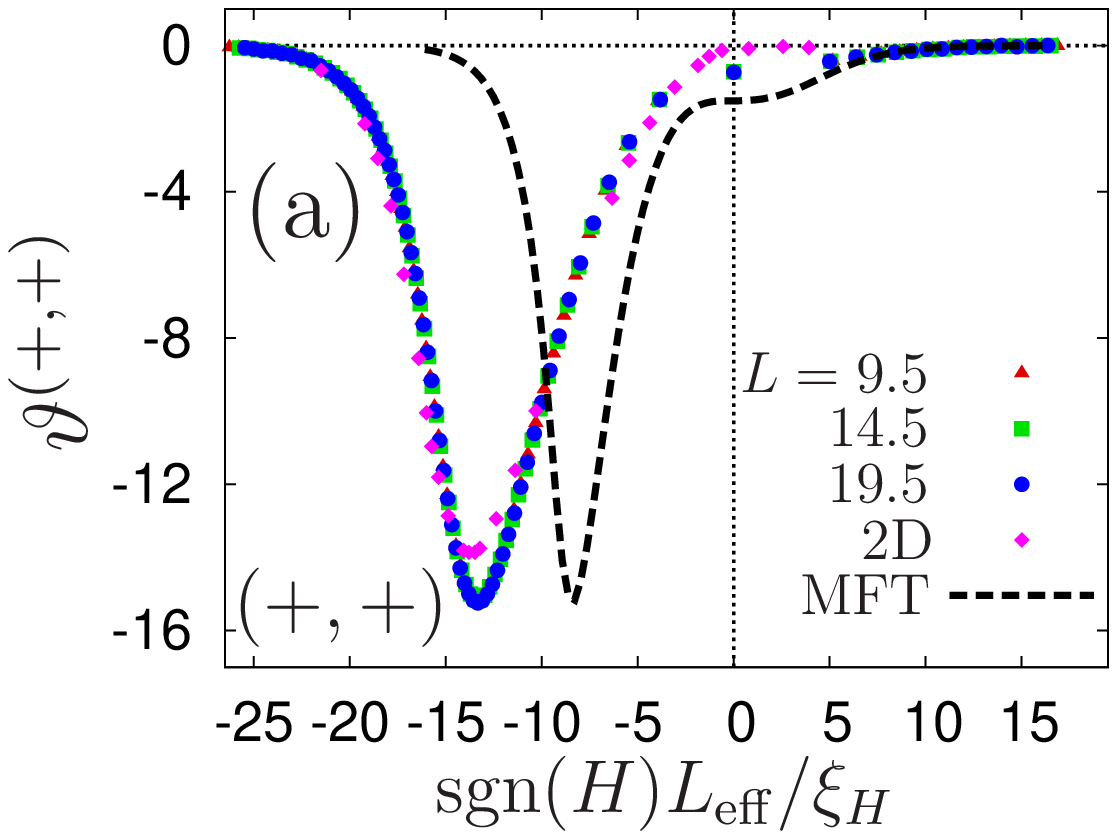}
\includegraphics[width=0.45\textwidth]{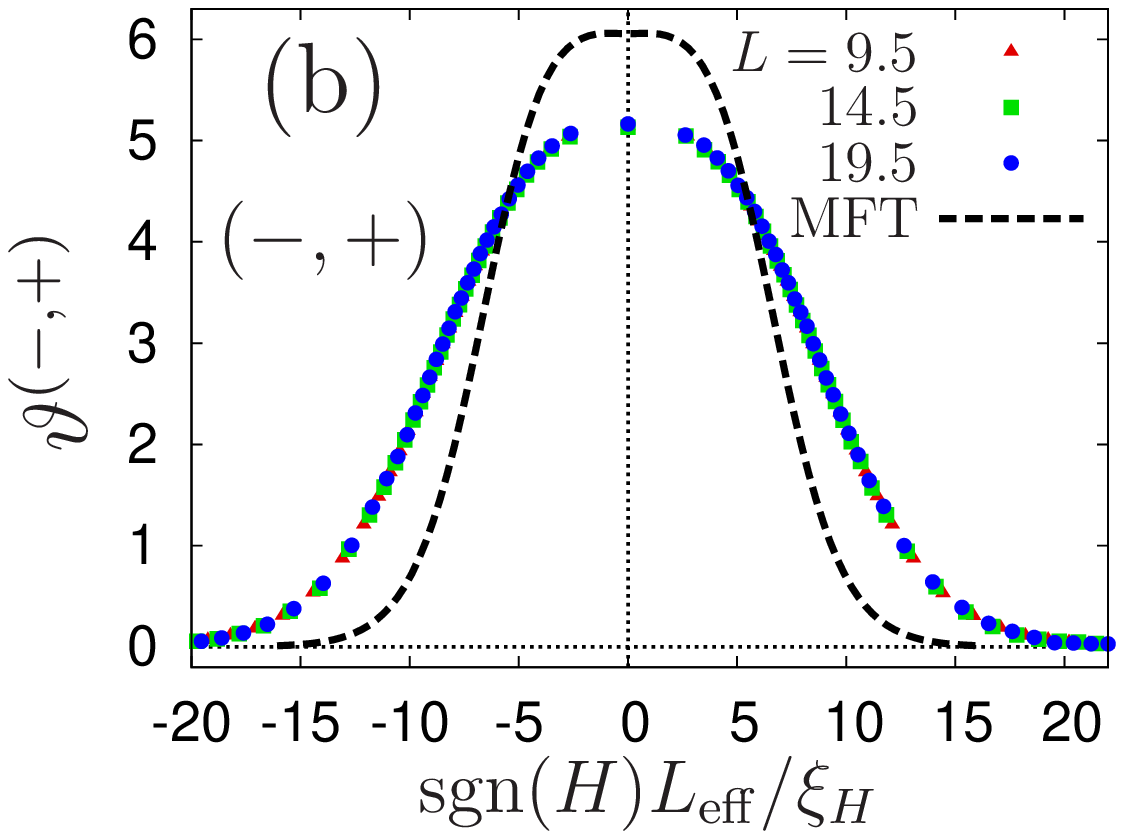}}\\
\mbox{\includegraphics[width=0.45\textwidth]{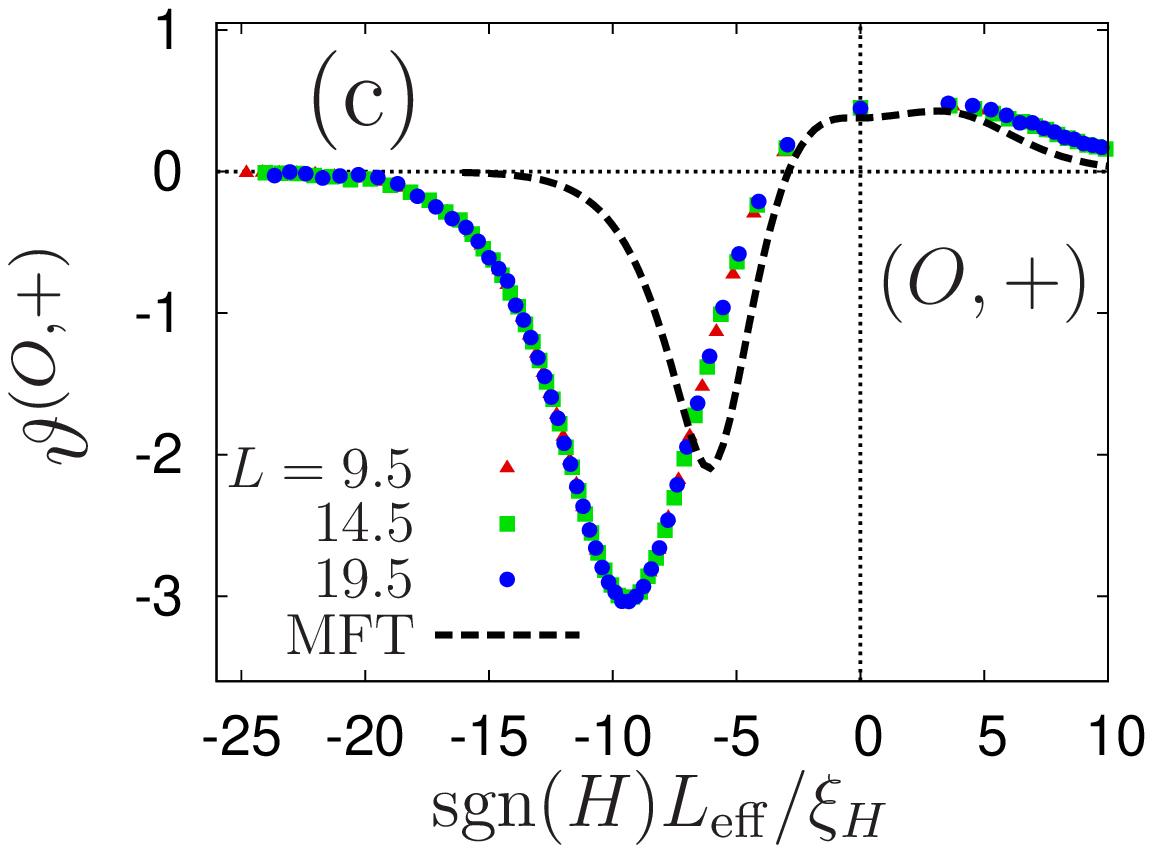}
\includegraphics[width=0.45\textwidth]{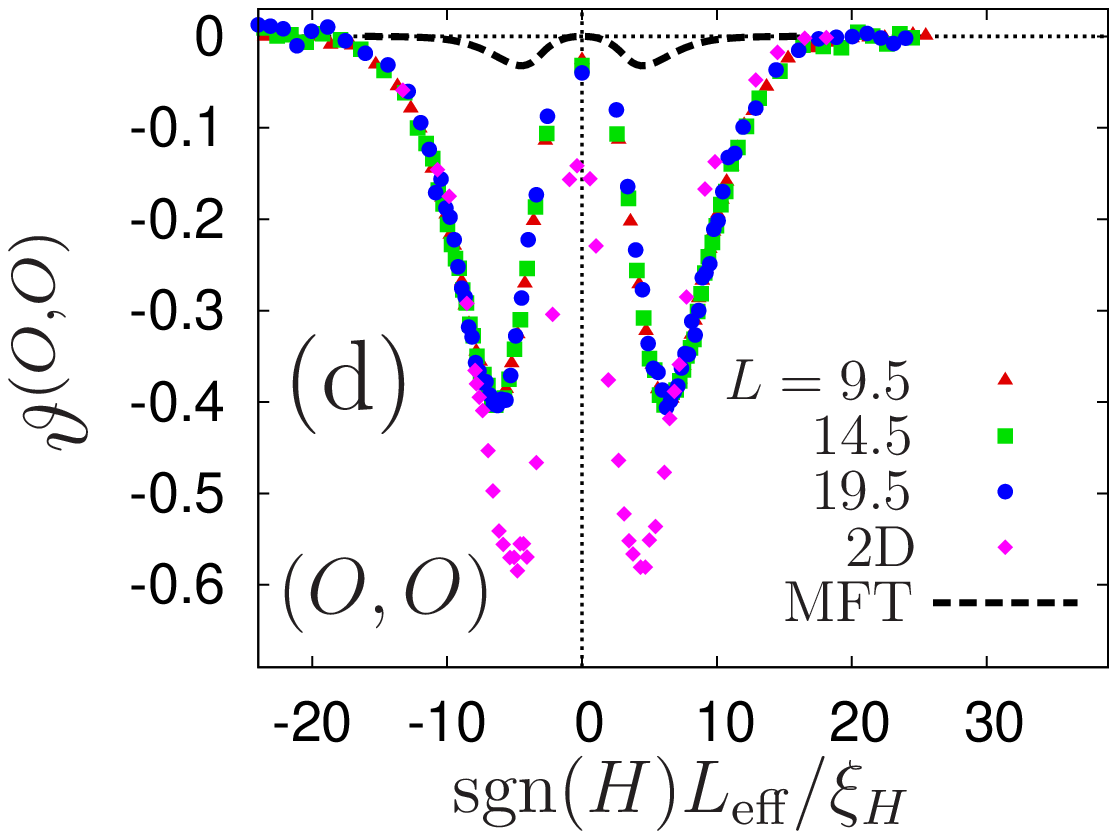}}
\caption{%
(a)-(d) show the same MC data as in Fig.~\ref{fig:thetad},
but not normalized. Here, the undetermined prefactor $g^{-1/2}$
of the MFT has has been fixed such that the depths of the minima in (a)
are the same. This value of $g \simeq 187.5$ has been used for the MFT results in (b)-(d).
For BC $(+,+)$ and $(O,O)$ (a) and (d) provide a comparison of the scaling
functions in $d=4$ and 3 with  those in $d=2$~\cite{MDC}.
The effect of stronger fluctuations in $d=2$ is most pronounced for free BC. 
In $d=2$ one has $\xi_{H,0}=0.233(1)$~\cite{ZMD}. 
}
\label{fig:theta}
\end{figure*}

On the basis  of  finite-size 
scaling theory~\cite{Barber,nature,PRE1,EPL,PRE,sp,Diehl}, in spatial dimension
$d$  CCF in units of $k_{\mathrm B}T$ and per $d-1$-dimensional area
 are expected to exhibit  the scaling form   
\be
\label{eq:sc}
\hat f_{\mathrm {C}}^{(BC)}(\beta,H,L)
=L_{\mathrm{eff}}^{-d}\vartheta^{(BC)}_{\pm}\left( L_{\mathrm{eff}}/\xi_{t}^{\pm},
L_{\mathrm{eff}}/\xi_{H}\right),
\ee
where the universal  scaling function $\vartheta^{(BC)}_{\pm}$ depends on 
 the boundary conditions at the  top and at the  bottom  surface, and
  $\xi_{H}=\xi_{H,0}|H|^{-\nu/\Delta}$ is the bulk correlation length at 
 $T=T_{c}$.
(Concerning the relationship between the scaling variable 
$h=H L_{\mathrm{eff}}^{\Delta/\nu}
 \sim \left(L_{\mathrm{eff}}/\xi_{H} \right)^{\Delta/\nu}$
and the physical quantity $(c_{A}-c_{A}^{c})/ c_{A}^{c}$ see Subsec.~II.B.1
in Ref.~\cite{PRE1},  Subsec.~II.B.2 and the Appendix in Ref.~\cite{Mohry1},
and Ref.~\cite{SHD}.)
 In eq.~(\ref{eq:sc}), for each BC we use an effective thickness
$L_{\mathrm{eff}}=L+\delta L$ such that, to a certain extent,
$\delta L$ captures some  corrections to scaling~\cite{Surf,Hass}.
Since here we are studying the  behavior of the CCF
{\it at} the critical temperature $\beta_{c}$,
one has $L_{\mathrm{eff}}/\xi_{t}^{\pm}=0$; thus in the following we
omit the first argument of 
 $\vartheta_{\pm}^{(BC)}(0,L_{\mathrm{eff}}/\xi_{H})
\equiv \vartheta^{(BC)}(L_{\mathrm{eff}}/\xi_{H})$.
We apply the  fitting procedure 
described in the  Appendix of Ref.~\cite{PRE}
which for each type of BC  minimizes the spread among the 
results for $\vartheta^{(BC)}$ as
obtained for various values of  $L\;(=9.5,14.5,19.5)$. 
This procedure  renders
the correction  $\delta L$ to scaling  (see Table~\ref{tab:dl}).
\begin{table}
\caption{
Correction $\delta L$ to scaling   for four BC.}
\begin{tabular}{|c|c|c|c|c|}
\hline
 (BC) & $(+,+)$ & $(-,+)$ & $(O,+)$ & $(O,O)$\\
\hline
 $\delta L$ & 0.60(10) & 0.65(2) & 0.93(10) & 1.22(2) \\
\hline
\end{tabular}
\label{tab:dl}
\end{table}
In Figs.~\ref{fig:thetad} and~\ref{fig:theta}
we plot the results for the CCF scaling function 
$\vartheta^{(BC)}$ as a function of the  scaling variable  
$\mathrm{sgn}(H)L_{\mathrm{eff}}/\xi_{H}$
 for the BC $(+,+)$, $(-,+)$, $(O,+)$, and $(O,O)$, respectively.
 Along the critical isotherm one has $\xi_{H}=\xi_{H,0}|H|^{-\nu/\Delta}$
where $\xi_{H,0}=0.3048(9)$ (see Ref.~\cite{EFS}).
After taking into account the aforementioned finite size corrections $\delta L^{(BC)}$,
for each BC we observe data collapse  
onto a master curve for different values of $L$.

It is instructive to compare these universal scaling functions  for $d=3$
with  those for $d=4$ which follow from minimizing the 
 Landau-Ginzburg Hamiltonian corresponding to eq.~(\ref{eq:Ham})~\cite{Diehl}:
\be
\begin{array}{c}
\mathscr{H}= \tilde A \int \limits_{0}^{\tilde L} \left[\frac{1}{2}
\left(\frac{{\mathrm d} \phi}{{\mathrm d}z  } \right)^{2}
+\frac{1}{2}\tau \phi^{2}+\frac{g}{4!}\phi^{4}
-\tilde H\phi \right] d z+\\+ \tilde A \left[
\frac{1}{2}c_{-}\phi_{0}^{2}-\tilde H_{1}^{-}\phi_{0}+
\frac{1}{2}c_{+}\phi_{1}^{2}-\tilde H_{1}^{+}\phi_{L}
 \right],
 \end{array}
\ee
where  $\exp\{-\mathscr{H}[\phi]\}$  is the statistical weight
of the scalar order parameter field, $\tilde A$ is the three-dimensional
cross-sectional area, $\tilde L$ is the film thickness, $g>0$,
 $\tilde H$ is the bulk field, $\tilde H_{1}^{-}$ and $\tilde H_{1}^{+}$
are bottom and top surface fields,
$\phi_{0}=\phi(z=0)$, and $\phi_{L}=\phi(z= \tilde L)$.
Within mean field theory (MFT), $1/c_{i}$ $[i=-,+]$ are extrapolation lengths~\cite{Diehl},
$\tau=(\xi_{t,0}^{+})^{-2}t $ for $t>0$, and $\tau=(\sqrt{2}\xi_{t,0}^{+})^{-2}t $
for $t<0$.
 Here and  below  we use  the tilde $\tilde{ \rule{0pt}{4pt}}$ 
 to mark  MFT quantities.
The solution of the corresponding Euler-Lagrange equation
renders the equilibrium order parameter profile:
\be
\frac{{\mathrm d}^{2}\phi}{ {\mathrm d}z^{2}}-\tau \phi(z)-
\frac{g}{6}\phi^{3}(z)+\tilde H=0
\ee
with the BC
\be
\left.\frac{{\mathrm d}\phi}{{\mathrm d} \tilde z}\right|_{\tilde z=0}
=c_{i}\phi(\tilde z=0)-\tilde H_{1}^{i}
\ee
where  $\tilde z$ is the separation from the wall, i.e.,
$\tilde z=z$ for $-$ and $\tilde z= \tilde L-z$ for $+$.
For large $\tilde H_{1}^{i}$, the leading behavior of $\phi(\tilde z \ll \tilde L)$
is given by $\pm \sqrt{12/g}{\tilde z}^{-1}$~\cite{SHD}  which corresponds to $\pm$ BC.
For large $c_{i}$ one has $\phi(\tilde z=0) \;=0$
corresponding to BC~O.

The  stress tensor  is
\be
T_{zz}(z,\tau,\tilde H)=
\frac{1}{2}
\left(
\frac{{\mathrm d}\phi}{ {\mathrm d}z}\right)^{2}-\frac{1}{2}\tau \phi^{2}(z)-
\frac{g}{4!}\phi^{4}(z)+\tilde H \phi(z),
\ee
so that the CCF in units of $\tilde A$ and $k_{\mathrm{B}}T$  equals
\be
f_{c}(\tau,\tilde H,\tilde L)=T_{zz}(z_{0},\tau,\tilde H)-T_{zz}^{b}(\tau,\tilde H)  
\ee
where $z_{0}$ is an arbitrary point $0 \le z_{0} \le \tilde L$.
The bulk contribution is
\be
T_{zz}^{b}(\tau,\tilde H)=
-\frac{1}{2}\tau \phi^{2}_{b}-
\frac{g}{4!}\phi^{4}_{b}+\tilde H \phi_{b}
\ee
with $\phi_{b}$ as the solution of
$\tau \phi_{b}+\frac{g}{6}\phi^{3}_{b}=\tilde H$.
For comparison, in  Figs.~\ref{fig:thetad}~and~\ref{fig:theta}  
we plot also the results for the normalized scaling functions
$\tilde \vartheta^{(BC)}/|\tilde \Delta_{+,+}|$
as a function of 
$\mathrm{sgn}(\tilde H)\tilde L/\xi_{\tilde H}$, 
where $\xi_{\tilde H}=\frac{1}{\sqrt{3}}|\tilde H|^{-1/3}$.
These functions describe the universal behavior in $d=4$.
 
 The CCF  for $(+,+)$ BC [see Figs.~\ref{fig:thetad}(a) and~\ref{fig:theta}(a)]
 is attractive.
 In $d=3$ the scaling function
  has a minimum at   $\mathrm{sgn}(H)L_{\mathrm{eff}}/\xi_{H} \simeq -13.3$
 for which the direction of the bulk field is opposite to that of the  surface fields.
 The depth of this minimum is   ca. 20.2 times 
  the value of the force at the critical point $(T_{c},H=0)$. 
 This means that the critical Casimir attraction between colloids
 suspended in a critical solvent 
  can be increased substantially by increasing 
  the concentration of that component of the binary liquid mixture 
  which is not preferentially adsorbed at the
   surfaces of the colloidal particles.
  For $(O,+)$ BC the force is attractive for strong,
 negative values of $H$ and   repulsive
 for  $H>0$ [see Figs.~\ref{fig:thetad}(c) and~\ref{fig:theta}(c)]. 
  The scaling functions for $(-,+)$ and $(O,O)$ BC  
 are symmetric with respect to $H=0$.
 For $(-,+)$ BC the scaling function has a maximum at 
 the critical point $H=0$ 
 [see Figs.~\ref{fig:thetad}(b) and~\ref{fig:theta}(b)]. The CCF for $(O,O)$
 BC is weakly attractive; the corresponding scaling function in $d=3$
 has two symmetric minima at  
 $\mathrm{sgn}(H)L_{\mathrm{eff}}/\xi_{H} \simeq  \pm 6.2$.
 
 In summary, we have carried out the energy 
integration method in order to compute
the bulk free energy density 
of the three-dimensional Ising model 
in the presence of a  bulk magnetic field $H$.
On this basis, by using a coupling parameter approach we have determined 
the  scaling functions of CCF for slabs of thickness $L$
  at the critical temperature $T=T_{\mathrm c}$
as a function  of the scaling variable
 $\mathrm{sgn}(H)L_{\mathrm{eff}}/\xi_{H}$.
The universal scaling  functions have been 
computed for the four types  $(+,+)$, 
$(-,+)$, $(O,+)$, $(O,O)$  of BC  and  have been 
compared  with results 
 in $d=4$ and in $d=2$ as far as available. 
At $T=T_{c}$, for all considered BC
except  $(-,+)$ the CCF attain their largest strength off two-phase
coexistence, i.e., for $H \ne 0$.

\end{document}